# Enhanced Coalbed Methane Extraction by Geothermal Stimulation in Deep Coal Mines: An Appraisal

Xu Yu[a, b, c], Baiquan Lin[a, b*], Cheng Zhai[a, b], Chuanjie Zhu[a, b], Klaus Regenauer-Lieb[c], Xiao Chen[c]

a. Key Laboratory of Coal Methane and Fire Control, China University of Mining and Technology, Xuzhou, China.

b. School of Safety Engineering, China University of Mining and Technology, 221116 Xuzhou, China.

c. School of Mineral and Energy Resources Engineering, UNSW Sydney, Australia.

**Abstract:** Coalbed methane embedded in coal seams, is an unconventional energy resource as well as a hazardous gas existing in mining industries, which attracts lots of global attention. As the largest coal producer, the mining industry in China had to deal with many hazards induced by methane for decades. To solve this issue, underground methane extraction is commonly used in underground coal mines. However, underground methane extraction is hampered by low production rate and low efficiency because of slow gas emission from coal primarily controlled by gas desorption and permeability. It is well known that temperature has a great impact on gas sorption. The higher the temperature the larger the desorption rate. As the depth of coal mines increases beyond 1000m coal mines suffer elevated air temperatures caused by the natural geothermal gradient. The elevated temperature in such mines provides a potential economical way for geothermal energy extraction and utilization in deep coal mines which can largely cut the expenses of installation and operation maintenance. Therefore, a novel method is proposed to enhance underground methane extraction by deep heat stimulation. This paper mainly presents an assessment of previous and ongoing research in the related field and provides a first feasibility analysis of this method applied in the underground environment. The technique proposed in this early appraisal is deemed significant for coalbed methane drainage enhancing the productivity of deep coal mines by geothermal technology and can also be extended for many applications in relevant areas such as shale gas, and tight oil.

---

[*] Corresponding author.
Email address: lbq21405@126.com

## 1 Introduction

Coalbed methane (CBM), also called coal mine gas (CMG), was first noticed hundreds of years ago as a dangerous factor in underground mines which has cost thousands of lives [1]. In recent decades, CBM has become an alternative energy resource for conventional fossil fuels that attracts the interest of many scientists and engineers from various fields including petroleum engineering, geoscience, and mining engineering [2, 3]. Currently, there are two types of methods for CBM recovery containing gas production on the surface with vertical wells and gas drainage with boreholes in underground mines. As 90% of coal mines in China are of shaft mining and 70% of coal seams are in low permeable and highly adsorptive strata the production of methane becomes a difficult technology [4]. The commonly used gas drainage techniques are of low production rate and quick decline because of low permeability and gas release rate [5-7]. The desorption rate increases with the increase of temperature thus providing a potential key to improve gas-production rate by heat injection. The geothermal energy sourced from surrounded strata commonly known as heat hazards in deep coal mines potentially provides an economical heat source. In this appraisal we assess this theoretically promising method in a desktop study to improve the emission rate of methane from coal and meantime, eliminate the effect of heat hazards on human's health.

Geothermal energy is a relatively sustainable and economical low-carbon energy resource buried in the earth [8]. According to existing data, the air temperature in many deep coal mines reaches 40 °C or above rendering the underground working conditions unbearable for miners. Previous research reported that high working temperature can seriously affect coal workers' health [9]. Deep coal mines are currently suffering from serious high-temperature issues (also called heat hazards) [10, 11]. To solve the heat hazards, the most common way is by providing a cooling system at high capital expense in excess of millions of dollars for its installation and operation maintenance [12]. However, the cost can be largely reduced by utilizing geothermal energy in underground space to stimulate CBM production. The extracted heat from surrounded hot strata is used to heat the CBM-production region, which can largely increase the methane release rate from the coal microstructures. However, geothermal extraction from underground coal mines with concurrent stimulation of CBM recovery has so far received little attention. In this contribution separately conducted pioneering work on geothermal extraction from abandoned mines and enhanced CBM by heat injection are reviewed.

Practical work on geothermal recovery from abandoned mines have been carried out for decades and shown to be a robust technology for Heating Ventilation and Air Conditioning (HVAC) in many countries [13-20]. These cooling and heating systems accomplished a large saving of energy consumption and economical expenses. Enhanced CBM by geothermal stimulation has so far only been discussed in theoretical and laboratory work on the sensitivity of methane sorption on coal by heat injection. The work by Boxho et al. shows that the methane adsorption volume decreases by 0.8% for every 1℃ increase in temperature [21] whereas Hofer et al. and Bustin et al. found a 2.2% and 1%-3% decrease in methane adsorption [22, 23]. The thermal effects on methane adsorption are not consistent on different adsorbents. But the results reveal the possibility and feasibility of the enhanced CBM technique by heat injection. Therefore, in recent years, new techniques have been proposed and practically applied to enhance the CBM recovery by heat injection with hot steam [15], hot air from power plants [16], or any other heat sources near coal mines have been proposed. Although heating coal seams can stimulate CBM extraction, the thermal stimulation technique has not found wide industrial usage because of the high costs for building thermal injection systems and their maintenance.

China, as the largest coal-production country, has many deep coal mines that are suffering from heat hazards caused by natural geothermal energy. Heat hazards include heat-related illnesses, injuries, and exhaustion occurring in the workplace caused by exposure to extreme heat. In the recent decade, the heat hazard is increasingly serious in coal mines with the increase in excavation depth. In the Mideastern basins of China, the mining depth already exceeds 1000m where the air temperature of the working face reaches 35 °C or higher. For instance, in the Pingdingshan coalfield, the peak temperature of the excavated strata has reached 46 °C with an average gradient of 3.2~4.6 °C/100m. The temperature can be much higher in some areas.

This type of temperature gradient is suitable for the installation of a geothermal extraction system in which water is circulated to bring the heat to the specific gas-recovery area.

This work summarizes the status of the research work in the related field and a feasibility analysis of the proposed technique. The main contents include the following sections. Section 2 presents the global development of the geothermal extraction from abandoned mines and the status of heat hazards in deep coal mines. Section 3 demonstrates the research work of the effects of temperature on CBM production and coal microstructures containing the temperature dependence of methane sorption, thermal conductivity, and permeability. In section 4, the feasibility of the utilization of deep geothermal energy to stimulate CBM extraction is discussed. Conclusions are presented in the following section 5.

## 2 Research status of geothermal heat extraction and heat hazards in deep coal mines

### 2.1 Global development of geothermal extraction from abandoned mines

An extremely large portion of geothermal energy is stored at an accessible depth in the earth's crust. Fig. 1 presents the distribution of global geothermal resources. To date, the total production of geothermal energy is very limited accounting for about 0.15% (0.565 EJ/year) of the total global energy consumption in 2015 [24, 25]. Underground mines were introduced as potential areas for geothermal extractions using hot mine water from the fractures and voids of the mining space after the abandoning mining operations[17]. Abandoned mines mainly have two advantages including good conditions for the installation of geothermal extraction systems and high permeability for water circulation. A heat pump is commonly used in geothermal projects to heat the circulated water to achieve a higher temperature. The geothermal projects can also be reversed to provide air-conditioning for cooling houses. The efficiency of heat pumps is quantified by the coefficient of performance (COP), which is the rate of heat produced by work supplied. The smaller the difference between the inlet and the delivered temperature, the higher the pump efficiency, meaning the less energy is needed to achieve the heating or cooling requirements.

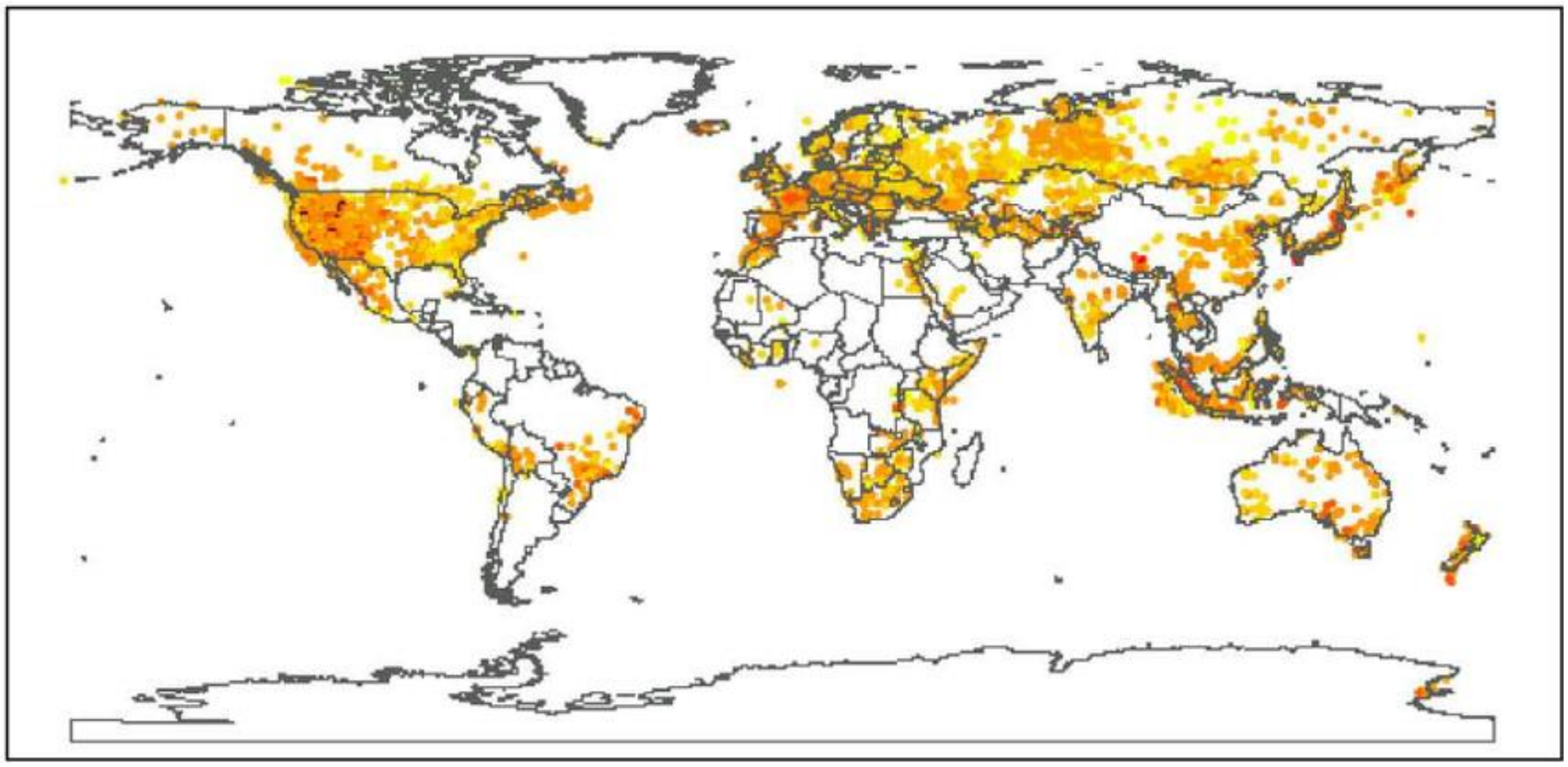

Fig. 1 Global distribution of geothermal resources referred to Pallack et al. 1993 [26]. The darkness of the yellow color represents the number of geothermal energy reserves.

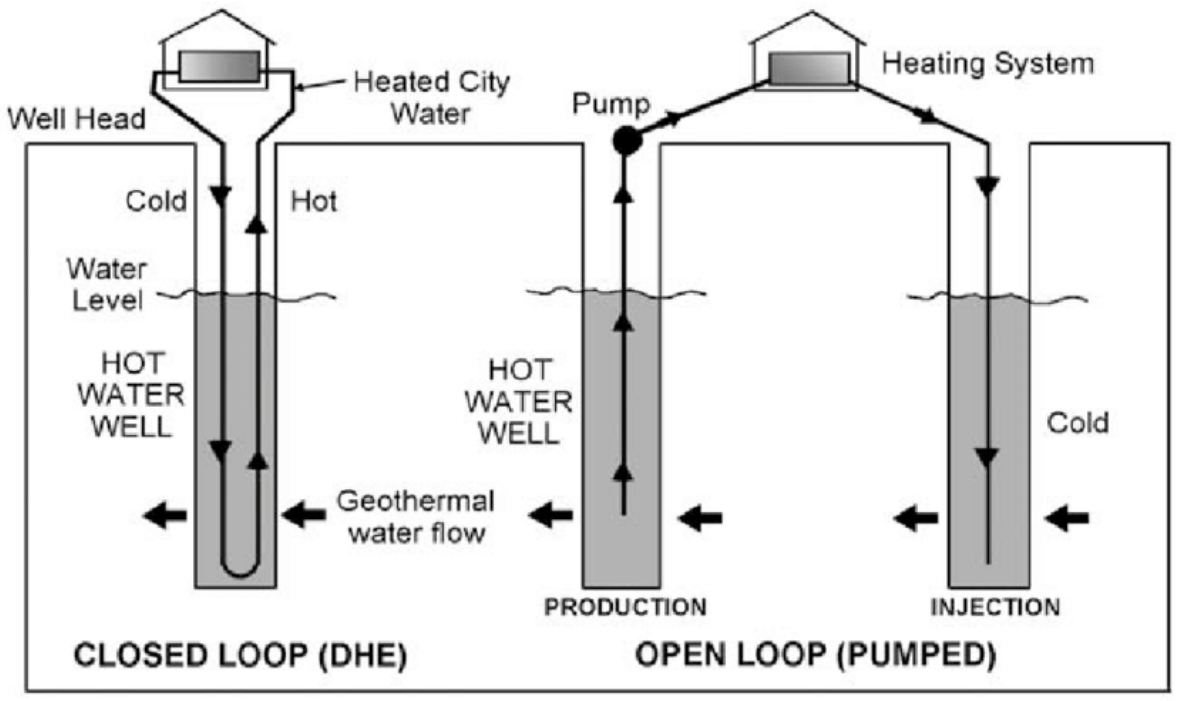

Fig. 2 Two typical geothermal systems including closed-loop (left) and open-loop (right) systems [27].

There are mainly two typical configurations for geothermal systems: open-loop or closed-loop systems [28]. An open-loop geothermal system, shown in the right part of Fig. 2, extracts the mine water and discharges it after capturing heat or releasing it. A closed-loop system, shown in the left part of Fig. 2, continuously circulates a heat transfer solution through buried or submerged plastic pipes in which the same solution is circulated in the closed-loop to provide heating and cooling with an indoor heat pump. In the existing cases of geothermal extraction from abandoned mines, the open-loop system is a more commonly used configuration and simpler than the closed system. A closed system usually requires a higher capital expense than an open system, due to the additional cost of materials such as pipes and drilling costs. But the closed-loop configuration can be used for mines with contamination issues or poor water circulation issues.

Many practical applications have been made all around the world including the US, Canada, UK, Germany, Netherland, and Spain presented in Table 1. The well-known geothermal extraction system based on an abandoned mine using hot mine water was installed in Spring Hill, Canada, in 1989 [17] which is designed to provide heating or cooling buildings. Due to the large reserve of geothermal resources in this region, another project was implemented in Quebec 20 years later [19]. Meanwhile, Germany has built several same geothermal systems mainly located in North-Rhine Westphalia, Saxony, and Anhalt. For example, the GrEEn project was implemented in the Eduard shaft in Alsdorf. The coal mine was filled with water with a temperature of about 26 °C after it was shut down in 1992. It is proposed to use hot mine water to heat the ENERGETICON energy museum. Another five geothermal systems were installed in Saxony, one in Bad Schlema, one in Ehrefriedersdorf, two in Freiberg, and one in Marienberg respectively [29]. Bad Schlema project was designed for an old school building for reducing the annual energy requirement while the geothermal systems in Ehrefriedersdorf, Freiberg, and Marienberg were for a middle school, castle, and mineralogical museum, University buildings, and commercial space respectively. The depths of those systems vary from 60m to 890m with a temperature range of 10-26 °C [18]. The Mine Water Project in Heerlen, located in the province of Limburg, Netherlands, is one of the largest geothermal district heating systems sourced by mine water. It aims to develop a sustainable full-scale smart grid to service 700,000 $m^2$ [30]. As the mine water is heavily polluted with sulphides, a closed-loop geothermal system located at the ‘‘Folldal Gamle Gruver’’ mining museum was applied in Folldal, Norway to heat the underground Wormshall chamber. Similar closed-loop systems were installed in Kongsberg, Norway, and Asturias, Spain. An assessment on the second project was conducted for the university buildings showing a total annual energy saving of 73% (1112050 kWh/year), a reduction of $CO_2$ emission of up to 39% per year, and monetary savings of 15% for the student residence and up to 20% for the research facility [31]. In the US, the practical implementation of geothermal recovery started early in the 1980s. A geothermal heating and cooling project based on heat pumps sourced by mine water has been carried out in 1981 in Kingston, Pennsylvania which has been working without complications to this day [32, 33]. Due to the success of the Kingston project, another two similar systems were built in Park Hills, Missouri (1995) and Scranton,

Pennsylvania (2010) aiming at the implementation of a cost-effective system that can be used for both building and process cooling. In the UK, two geothermal systems for space heating using the low enthalpy of the mine water were successfully installed in Shettleston, Glasgow, and Lumphinnans, Fife, Scotland which achieved annual savings of about 80% of on heating costs [13]. Table 1 shows that 80% of the 18 geothermal projects are configured with open-loop system. All the projects are used for heating or cooling the surface houses or buildings and contribute large annual energy and expense savings.

Table 1 A summary of the information on geothermal extraction projects under operation

| Location | Configuration | Depth/m | Temperature/℃ | Flow rate /m3/h | Heating capacity /KW | COP | End-user | Reference |
|---|---|---|---|---|---|---|---|---|
| Pennsylvania, USA | Open-loop | 91 | 15 | 20.2 | -- | -- | Building | Schubert et al., 1982 [17] |
| Spring Hill, Canada | Open-loop | 1350 | 26 | 41.7 | 3.73 | -- | Building | Jessop et al., 1995 [17] |
| Quebec Spring Hill, Canada | Open-loop | -- | 6.7 | 176.4 | 765 | 3.6-5.3 | Apartment building | Raymond et al., 2008 [19] |
| Alsdorf, Germany | Closed-loop | 890 | 26 | -- | -- | -- | Office buildings | Energeticon, 2013 |
| Freiberg (Castle), Germany | Open-loop | 60 | 10.2 | 10.8 | 130 | 3.5 | Castle and mineralogical museum | Kranz and Dillenardt, 2010 [29] |
| Freiberg (University), Germany | Open-loop | 216 | 18 | -- | 260 | 4.0 | University building | Grab et al., 2010 [18] |
| Marienberg, Germany | Open-loop & Closed-loop | 105 | 12.4 | 120 | 310 | 5.6 | Two commercial installations, tennis hall and adventure pool | Matthes and Schreyer 2007 [18] |
| Wettelrode, Germany | Open-loop | 283 | 12-13 | 90–150 | 47 | -- | Mining museum | Koch and Hoffmann, 2013 [18] |
| Heerlen, Netherlands | Open-loop | 700 | 27-32 | 120-230 | -- | 5.6 | Offices buildings and university | Verhoeven et al., 2013 [30] |
| Folldal, Norway | Closed-loop | 50 | 6 | -- | 18 | -- | Worm shall chamber (Cavern) | Kristoffersen et al., 2014 [18] |
| Rostov, Russia | Open-loop | 50–150 | 12–13 | -- | 40,000 | 3.5 | Different building of 3 areas | Ramos et al., 2006 [18] |
| Asturias, Spain | Closed-loop | 250 | 17-23 | 400 | 1000 | 5.5 | Research center and student residence of the university | Jardón et al., 2013[31] |
| Shettleston, UK | Open-loop | 100 | 12 | -- | -- | -- | Building (16 houses) | John Gilbert Architects, 2014 [18] |
| Lumphinnans, UK | Open-loop | 170 | 14.5 | -- | -- | -- | Houses (18) | Watzlaf and Ackman, 2006 [13] |
| Kingston, USA | Open-loop | 76 | 16 | 20.5 | -- | -- | Recreation venter | Schubert and McDaniel, 1982 [32] |
| Park Hills, USA | Open-loop | 153 | 13.9 | 273.6 | 112.5 | -- | Municipal building | Geothermal Heat Pump Consortium, 1997 [18] |
| Scranton, USA | Open-loop | 122 | -- | -- | -- | -- | University building | Korb et al., 2012 [33] |
| Xiaozhuang mine, Shanxi, China | Open-loop | 427 | 21-24 | 756 | 21980 | -- | Mine buildings | Zhang et al., 2019 [34] |

China also has a large reserve of geothermal resources in underground coal mines. As the excavation depth of the mining industry gets increases beyond 1000 m from the surface, the increased temperature affects the underground working conditions for shaft mining.

However, compared to the development of geothermal recovery in the above countries, the exploration and utilization of geothermal energy from deep coal mines have been underutilized in China. A geothermal system was installed in Xiaozhuang mine, Shanxi, China, for heating the buildings on the surface [34]. In addition, efficient solutions for eliminating the heat hazard in working space are needed for mining industries. Heat hazards in deep coal mines will be discussed in the next subsection.

**2.2 Status of heat hazard in deep coal mines**

The work by Stephenson et al. shows that many mines around the world are hampered by heat hazards [10]. The most well-known temperature hazard is found in the South African goldfields that go extremely deep. The geothermal gradient is relatively low around 12°C per kilometer, however, due to the great depth (4000m) and high surface temperature rock temperature of 60°C are not uncommon [35, 36]. Heat hazard is a serious issue faced by scientists and engineers from the mining industry. China is the largest coal-production and coal-consumption country in the world [37]. Coal is the primary energy source accounting for about 60% of the total annual energy consumption of China [38]. Therefore, the heat hazard, as one of the primary hazards, is encountered more frequently in Chinese coal mines of China than in other countries [12]. This section mostly introduces the heat-hazard status of coal mines in China. Due to the pace of exploitation of shallow coal reserves, the mining depth is increasing with time and becomes a growing causing a considerable increase in underground air temperature encountered by miners [11]. It is found that the workers' health is threatened by prolonged exposure to heat and humidity [39, 40]. The work productivity shows as a non-linear function with air temperature as shown by Koehn and Brown [41]. This issue has negatively affected the productivity as shown by the predictive model of Thomas and Yiakoumis [42]. Recently, Ibbs and Sunn developed a combined model based on Grimm and Wagner (G&W) [43] and Thomas and Yiakoumis (T&Y) studies [44]. Fig. 7 presents the results of a comparative study of the combined model, T&Y model, and G&W model. The combined model is defined as:

$$\text{Estimated productivity ratio}=23.92+2.07T-0.014T^2 \qquad (1)$$

where $T$ is the temperature in ℉. The high temperature in underground spaces causes great difficulty in creating and maintaining acceptable working conditions [45]. Therefore, rigorous cooling demands are required and are presenting greater challenges as a result of deeper mining activities and increased surface temperatures [46-48].

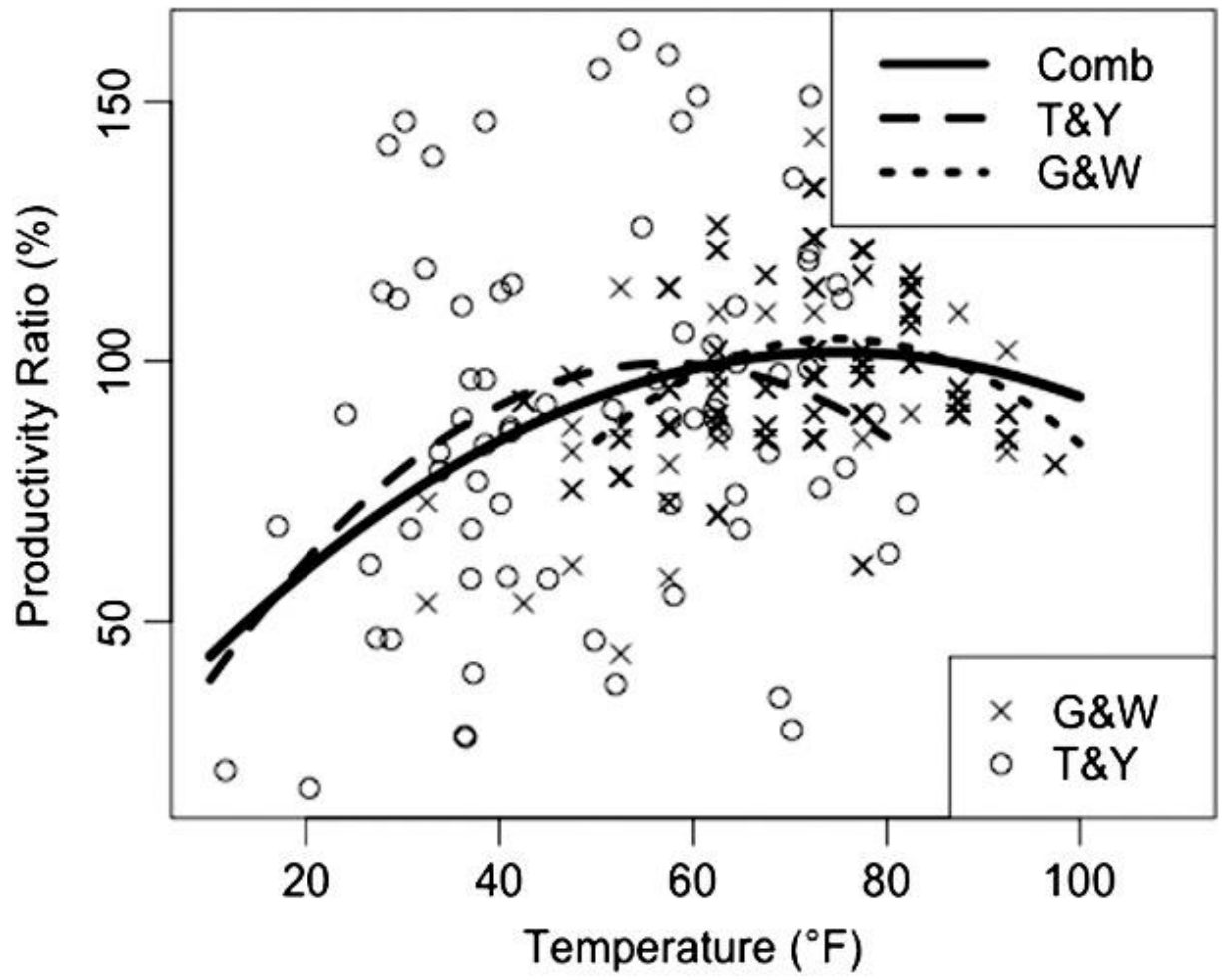

Fig. 3 Temperature versus productivity ratio with regression models referred to in the work by William Ibbs [44]. The solid line is a regression line based on combined data, the dotted and dashed lines represent the curves of T&Y and G&W models, respectively.

The work by Yang et al. shows that coal production in China has been suffering from heat hazards [12]. Fig. 4 and Fig. 5 present the distribution of major hydrocarbon basins, heat flow, and geothermal resources evaluation scope in China. It can be seen that the heat flows in the corresponding coalfields including Songliao, Ordos, Yellow sea, and Bohai basins, are about 60-90 mW/m$^2$. Table 2 gives a review of the high-temperature coal mines in the above coalfields including parameters of the mine location, excavation depth, temperature gradient, virgin-rock, and working-face temperatures. It shows that the mining depth in China has exceeded 1300 m with high heat in the virginal strata. The working-face temperature can reach as high as about 40 °C. From the results in Fig. 3, the production ratio decreases with the temperature increasing over 25 °C. Prolonged exposure to heat can cause disturbance of the body temperature of miners which may cause loss of concentration, heat illnesses, and, in extreme cases, death [43, 49]. Therefore, it is of great importance to manage the working temperature in the deep coal mine to protect workers’ health as well as mining productivity.

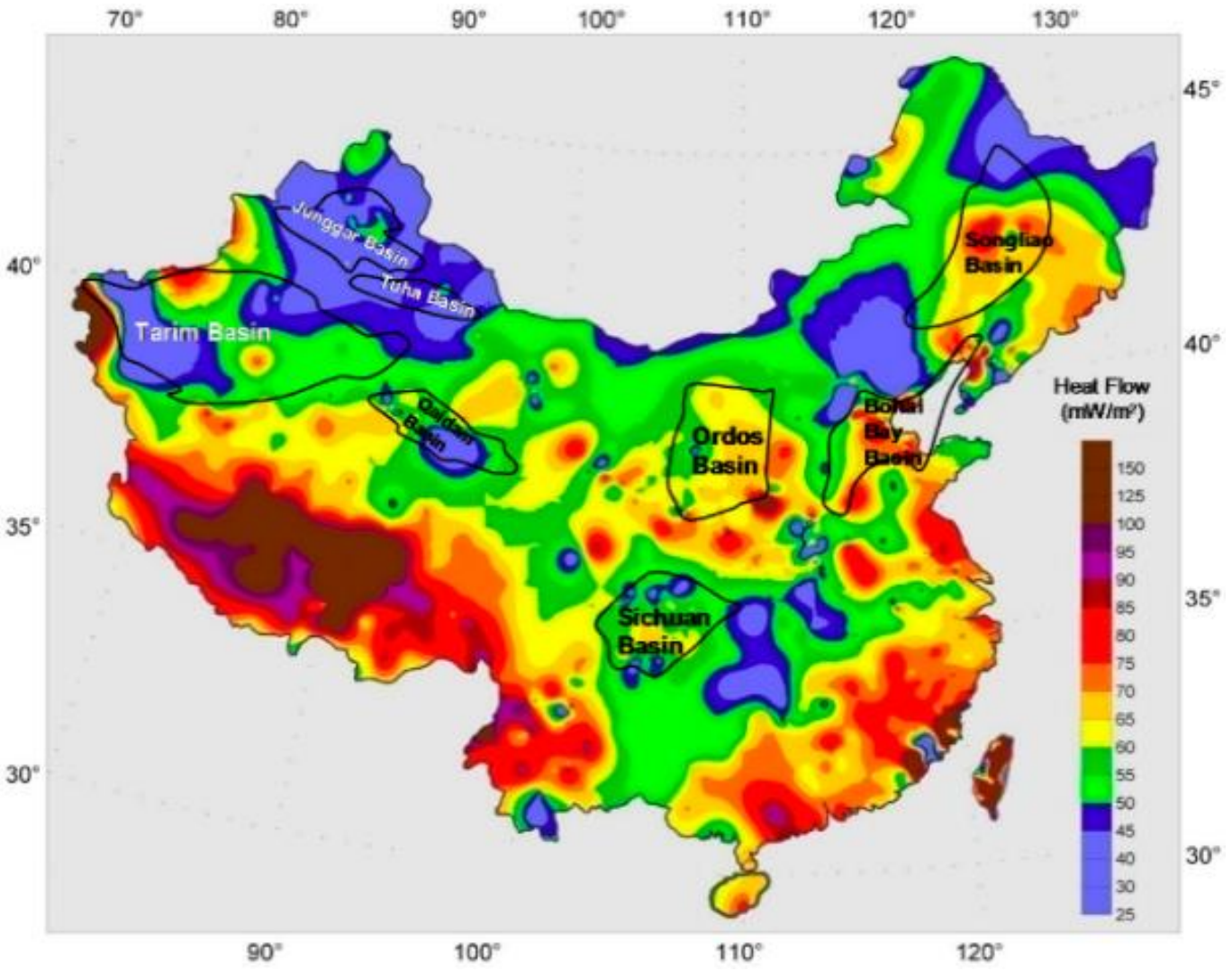

Fig. 4 Distribution of petroliferous basins and heat flow in China [50].

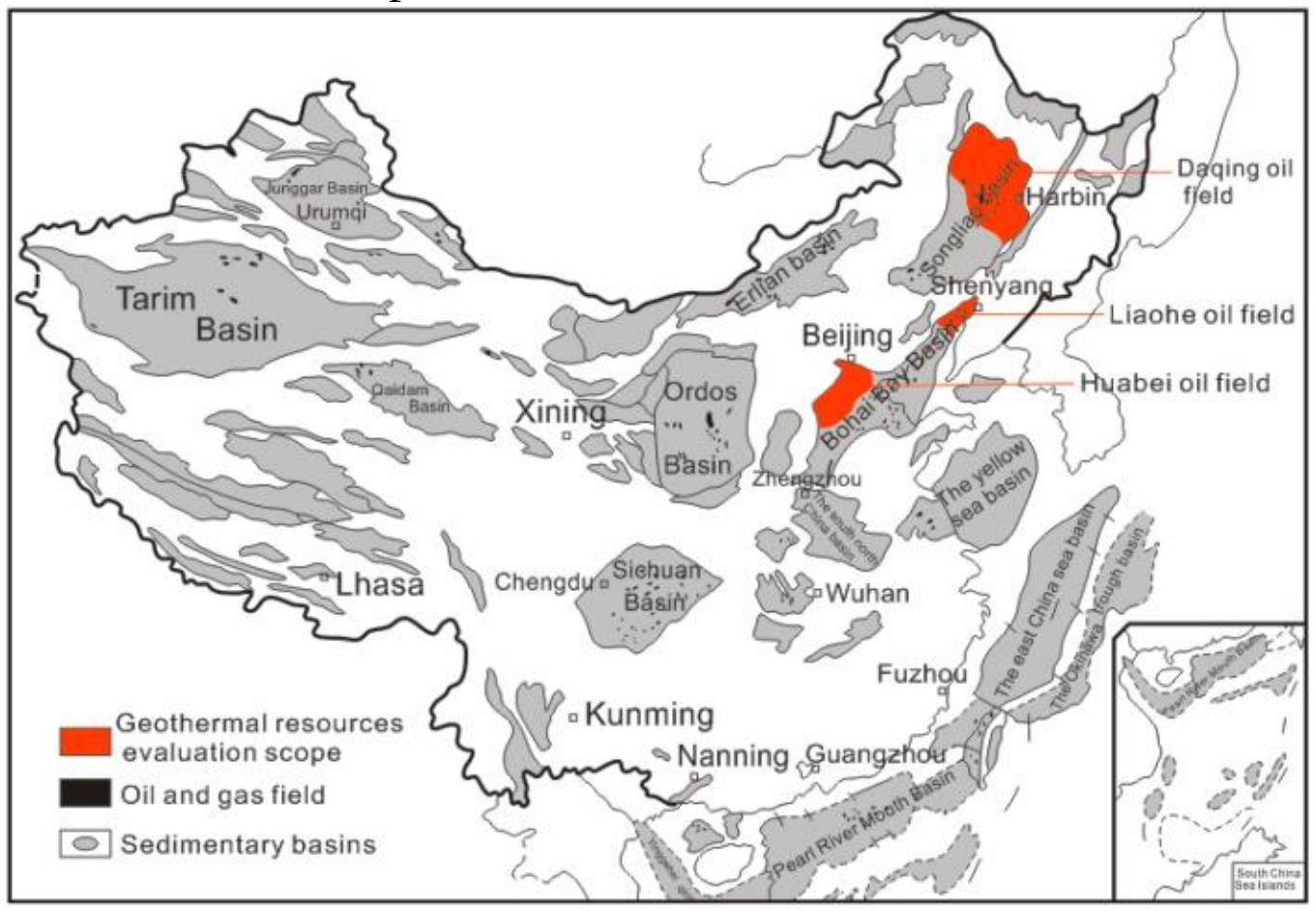

Fig. 5 Distribution of major hydrocarbon basins and geothermal resource evaluation scope [51].

Table 2 Summary of the parameters of deep coal mines suffered from heat hazards in China.

| Coal mines | Location | Depth /m | Temperature gradient /°C /100m | Virgin-rock temperature /°C | Working-face temperature /°C |
|---|---|---|---|---|---|
| Zhangxiaolou | Xuzhou, Jiangsu | 1125 | 2.52 | 44.8 | 36 |
| Sanhejian | Xuzhou, Jiangsu | 1300 | 3.32 | 50 | 39.6 |
| Jiahe | Xuzhou, Jiangsu | 1200 | 2.21 | 37.5 | 32-34 |
| Zhangshuanglou | Xuzhou, Jiangsu | 1000 | 2.60 | 43.7 | 34-36 |
| Mengchun | Xianyang, Shanxi | 700-800 | 3.76 | 16-27 | 30 |
| Gaojiabao | Xianyang, Shanxi | 980 | 3.18 | 46 | 25-30.8 |
| Yongchuan | Chongqing | 800 | 2.40 | 35-36 | 29.3-31.5 |
| Jining 2nd mine | Yanzhou, Shandong | 700 | 2.44 | 38-42 | 30 |
| Shangzhuang | Fengcheng, Jiangxi | 900 | 2.48 | 39.7 | 33-38 |
| Zhouyuanshan | Chenzhou, Hunan | 650 | -- | 38.4 | 30 |
| Fenglong | Xinyang, Henan | 900 | -- | -- | 31 |
| XiaoZhuang | Binzhou, Shanxi | 427 | 1.68 | -- | 34 |
| Pingmei 5th mine | Pingdingshan, Henan | 909 | -- | 42.6-46 | 33-35 |
| Jidong | Hebei | 730 | -- | -- | 28-31 |

## 3 Temperature dependence of CBM production

### 3.1 Effects of temperature on methane ad/desorption

Gas adsorption is primarily described by the Langmuir adsorption isotherm which shows adsorption for an ideal gas at isothermal conditions. Adsorption and desorption are reversible processes in this model which is treated as a chemical reaction as shown in Eq. (2).

$$A_{gas} + S \rightleftharpoons A_{adsorption} \tag{2}$$

Eq.(1) indicates that the adsorbate gaseous molecule $A_{gas}$ is adsorbed onto an empty sorption site, $S$, to yield an adsorbed species, $A_{adsorption}$. The Langmuir isotherm encapsulates the effect of pressure related to the volume of gas adsorbed onto a solid adsorbent at a given temperature. From these considerations, the mathematical formulation of the Langmuir adsorption isotherm can be derived and written as

$$\theta_A = \frac{V}{V_m} = \frac{K_{eq}^{A} p_A}{1 + K_{eq}^{A} p_A}, \quad (T\text{=constant}) \tag{3}$$

and can be rewritten as

$$\theta_A = \frac{P}{P + P_L} \tag{4}$$

where the temperature is assumed as constant, $\theta_A$ is the fractional occupancy of the adsorption sites defined by the ratio of the adsorbed gas volume $V$ to the maximum adsorption volume $V_m$, $K_{eq}^{A}$ is the equilibrium constant, and $p_A$ represents the gas pressure.

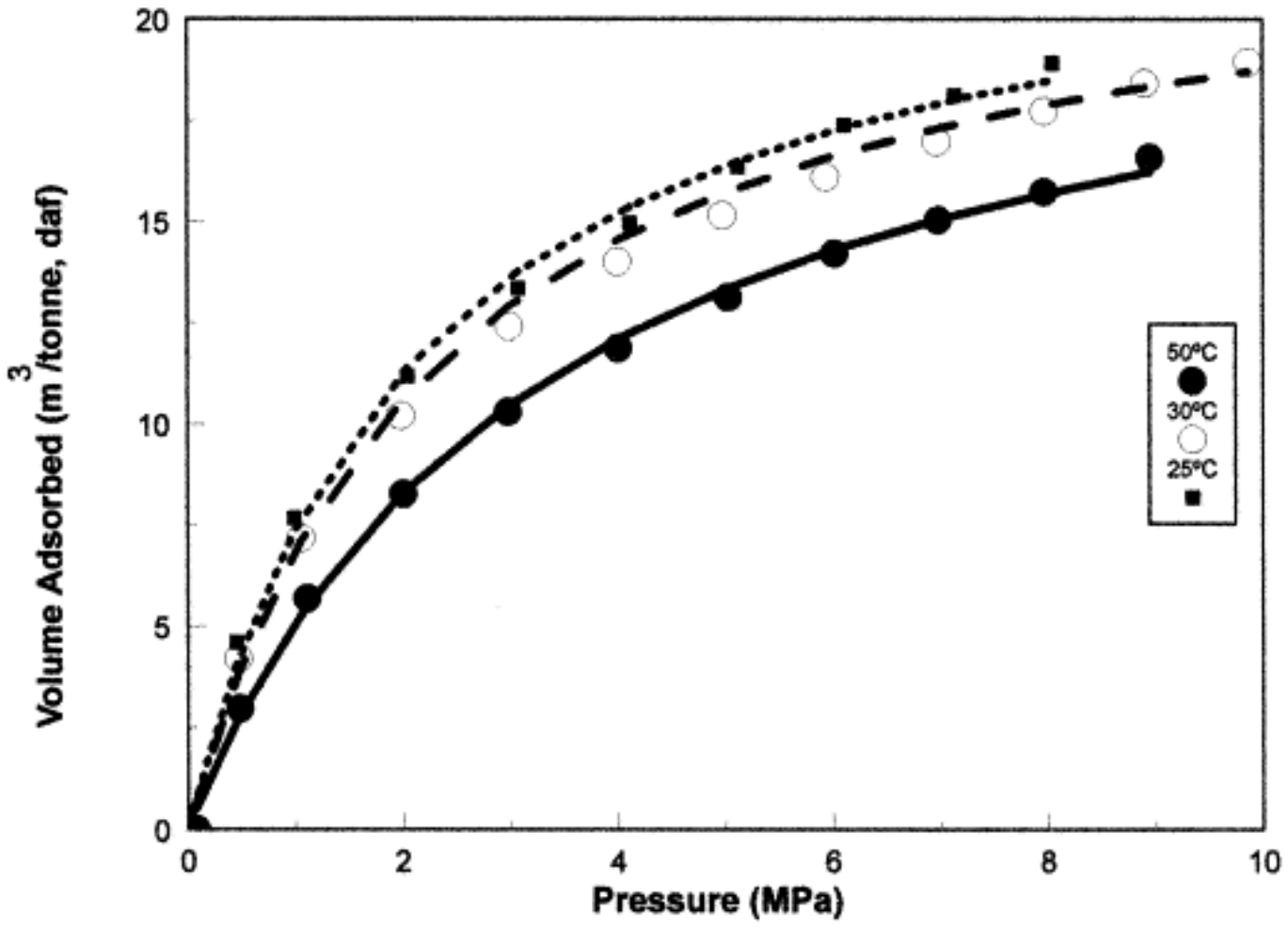

Fig. 6 Variation in methane adsorption with temperature for bituminous coal from the Bowen Basin, US [23].

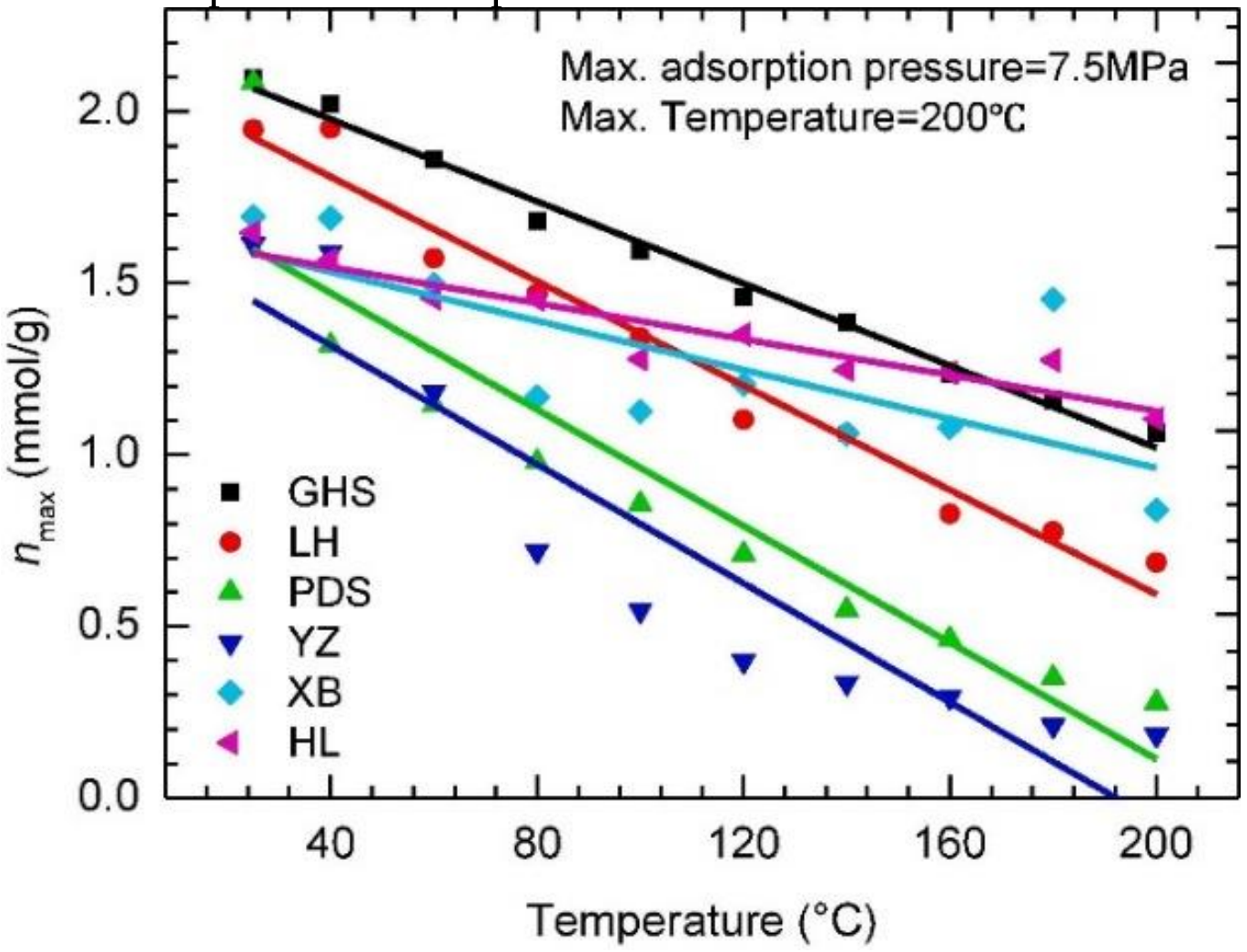

Fig. 7 Evolution of methane adsorption capacity along with the temperature sourced from Zhu et al. [52]. GHS-Anthracite, LH-Semi-anthracite, PDS-Medium volatile bituminous, YZ-High volatile A bituminous, XB-High volatile B bituminous, HL-Sub-bituminous.

The results of Bustin et al. in Fig. 6 shows that temperature plays an important role in the gas ad/desorption. The adsorbed volume of methane decreases by about 20% along with the temperature increasing from 25 ℃ to 50 ℃. A study of methane adsorption on different-rank coals under temperature and pressure conducted by our group shows that the methane adsorption capacity has a sharp decrease with the increase of temperature for all six coal samples [52]. As shown in Fig. 7, the adsorption capacities of all samples decrease with increased temperature, but it varies for different types of coal. The adsorption potential mentioned in this work defines an equilibrium between the chemical potential of gas near the surface and the chemical potential of the gas from a large distance away. The lower the temperature the smaller the adsorption potential, which indicates that more gas can be easily adsorbed on coal surface. The results show a good agreement with the previous literature data. A review of the related work on the dependence of the temperature of methane adsorption on coal is presented in Table 3. The results of the above work provide a basis for enhanced CBM techniques by increasing the temperature of coal seams such as heat injection.

Table 3 A review of the previous research on methane adsorption on coal at various temperatures [52].

| Coal sample | Coal type and sampling location | Pressure /MPa | T /℃ | FC /% | $R_o$ /% | VM /% | Reference |
|---|---|---|---|---|---|---|---|
| JB-1 | Semi-bright coal, China | 0-4 | 20, 50 | 42.8 | 0.62 | 40.1 | Cai et al., (2014) [53] |
| OB-1 | Semi-dull coal, China | | | 82.0 | 1.2 | 4.8 | |
| SQB-1 | | | | 74.8 | 1.89 | 1.4 | |
| Kyungdong coal | Anthracite, South Korea | 0-13 | 45, 65 | 57.39 | 5.07 | 5.01 | Kim et al., (2011) [54] |
| PB coke | Coke, Pittsburgh, US | 0-6.9 | 22, 102, 207, 307 | -- | -- | -- | Yang et al., (1985) [55] |
| PB coal | Bituminous, Pittsburgh, US | | | 52.7 | -- | 39.2 | |
| ML char | Lignite char, Montana, US | | | 51.0 | -- | 37.4 | |
| ML coal | Lignite coal, Montana, US | | | -- | -- | -- | |
| Jincheng mine | Semi-bright, China | 0-11.2 | 25, 35, 45 | -- | 2.74 | 5.42 | Wu et al., (2016) [56] |
| Coal A | Bowen Basin coal, Australia | 0–16 | 40, 50, 60, 70 | 63.5 | -- | 25.0 | Bae and Bhatia, (2006) [57] |
| Coal B | | | | 56.9 | -- | 26.3 | |
| Xingq-1-5# | Bituminous, Qinghai, China | 0-8 | 30, 40, 50 | 71.17 | -- | 29.83 | Hao et al., (2014) [58] |
| Qingh-2-3# | | | | 78.53 | -- | 21.47 | |
| Xujd-1# | | | | 88.34 | -- | 11.66 | |
| Leiy-1# | Anthracite, Hunan, China | | | 94.45 | -- | 18.45 | |
| Datong | Anthracite, Shanxi, China | 0-18 | 21, 38, 60, 80 | 51.18 | 3.72 | 10.48 | Zhang et al., (2015) [59] |
| GHS | Anthracite | 0-8 | 25, 40, 60, 80, 100, 120, 140, 160, 180, 200 | 83.35 | 3.09 | 8.63 | Zhu et al., (2019) [52] |
| LH | Semi-anthracite | | | 83.79 | 2.41 | 11.04 | |
| PDS | Medium volatile bituminous | | | 45.62 | 1.22 | 16.43 | |
| YZ | High volatile A bituminous | | | 53.97 | 0.84 | 12.42 | |
| XB | High volatile B bituminous | | | 57.75 | 0.58 | 11.34 | |
| HL | Sub-bituminous | | | 57.79 | 0.49 | 10.60 | |
| RU1 | Sub-bituminous, Huntly Coalfield, NewZealand | 0-8 | 25, 32, 40, 50, 60 | 42 | -- | 37 | Crosdale et al., (2008) [60] |
| J10 | | | | | -- | 37 | |
| Permian coal | Bowen Basin, Australia | 0-5 | 10-70 | 60-90 | -- | -- | Levy et al., (1997) [61] |
| Illinois | Illinois Basin, US | 0-8.27 | 10, 20, 30, 40, 50, 60, 70 | -- | -- | -- | Guan et al., (2018) [62] |

*Note: **P**-pressure, **T**-temperature, **FC**-fixed carbon, **$R_o$**-Vitrinite reflectance, **VM**-Volatile matter

### 3.2 Effects of temperature on the thermal conductivity and microstructures of coal

Heat exchange between geothermal fluid and coal is a key process in geothermal energy recovery. The thermal conductivity of coal largely controls the efficiency of the heat exchange. Thermophysical properties of coal are related to mineral composition, rank, temperature, volatile matter, compaction (and in consequence porosity), and anisotropy. Dindi et al. conducted a study for the measurement of the thermal conductivity and diffusivity of wet and dry coal samples from the Illinois Basin along with the temperature using the transient-hot-wire and transient-hot-plate methods [63]. The same method was used by Badzioch et al. to measure the thermal conductivity of 12 types of coal by lab experiments [64]. It is found that the thermal conductivity of coal samples changes in a small range with a temperature of 373-573 K (the mean value is about 0.23 $W\cdot m^{-1}$). For the temperature exceeding 573 K, thermal conductivity and diffusivity have a rapid increase. Turian et al. measured the thermophysical property of two types of coal (Illinois Ziegler mine no. 5 coal and Pittsburgh seam no.8 coal) at room temperature. The derived thermal diffusivity ($a = \frac{\lambda}{\rho C_p}$, defined by the measured values of $\lambda$, $\rho$, and $C_p$), were between 0.174 and 0.142 $mm^2\cdot s^{-1}$ [65]. The work by Ramazanova et al. shows the measured thermal conductivity varies from 0.341 to 0.497 $W\cdot m^{-1}\cdot K^{-1}$ for wet coal samples before thermal treatment and from 0.272 to 0.316 $W\cdot m^{-1}\cdot K^{-1}$ for dry samples after thermal treatment. The conductivity of wet and dry coal samples increases along with temperature reaching 390 K, and then gradually decreases for higher temperature.

Another study by our group is conducted to observe the effect of the heat on the coal microstructures [66]. Microwaves are used as a heat source for heating coal samples. The data indicates that microwave heating can considerably enlarge the micropores and enhance the connectivity of pores and fractures of coal. It can be seen from Fig. 8, that with the prolonged microwave heating, the $T_2$ amplitude largely changes which means that the coal microstructures have been modified. The change of the $T_2$ curve between before and after the microwave heating illustrates the modification of porous structures in coal. The number of micro-pores decreases for $T_2$ <0.8 ms and that of large pores increases for $T_2$ >0.8 ms. Therefore, the heating procedure has a positive impact on increasing the size of micropores and pore connectivity to achieve a higher permeability which can be used to enhance the CBM production at the macroscopic scale.

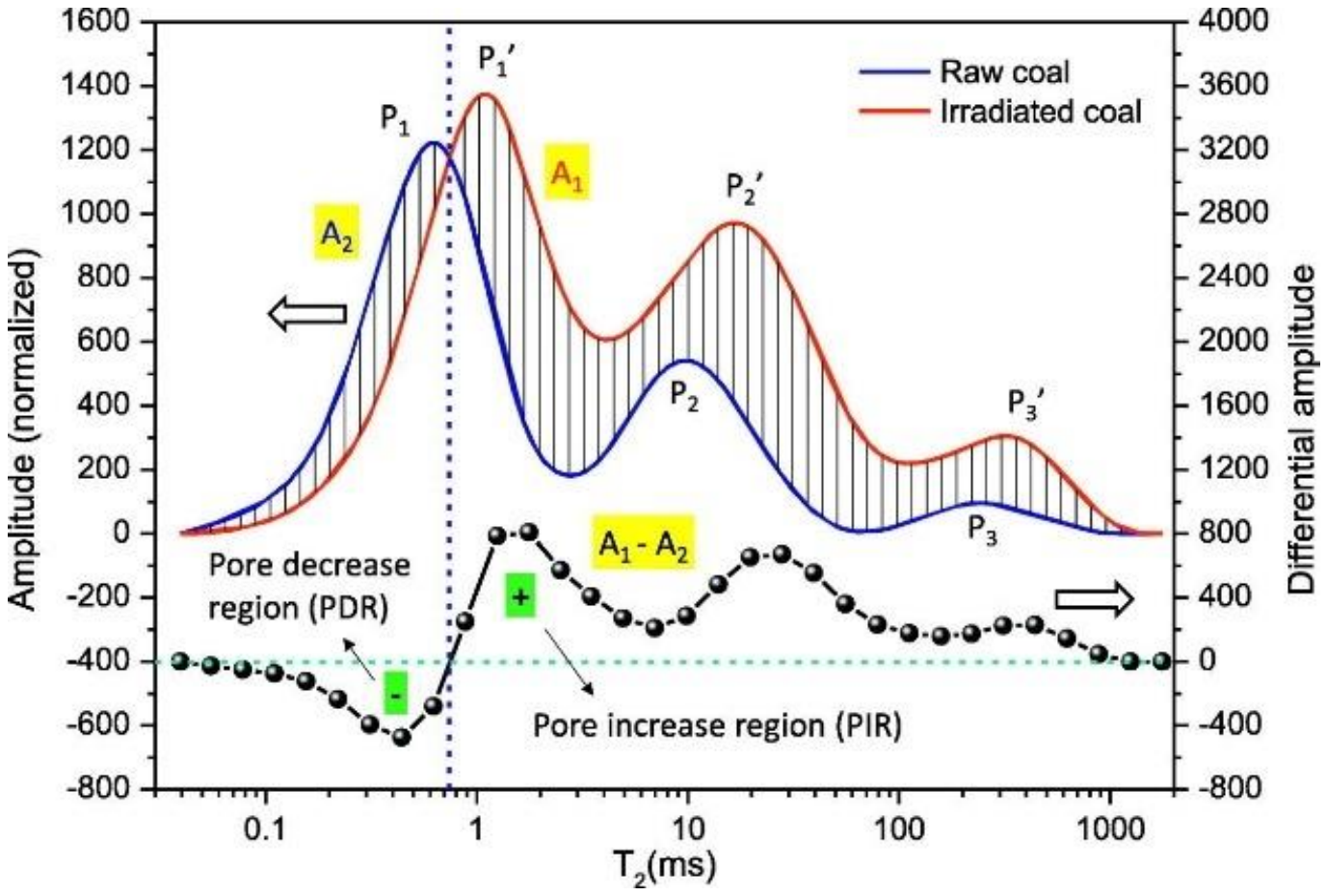

Fig. 8 Distributions of the $T_2$ spectra of the NMR results for the raw and heated coals [66]. The $T_2$ curve represents the distribution of varied-size pores in coal.

## 4 Discussion: feasibility analysis of enhanced CBM by geothermal stimulation

This paper introduces a novel enhanced CBM technique using geothermal stimulation which utilizes the geothermal energy extracted from the high-temperature region (e.g. a deeper aquifer) for heating the CBM drainage region. In this system, the thermal-extraction fluid is circulated through pipes and heated by a heat pump to reach a high enough temperature for heating the corresponding area as shown in Fig. 9. In section 2, we have shown that there are rich geothermal resources in underground coal mines. Many countries have practically used the mine water to extract the heat out of abandoned coal mines and saved a huge volume of energy and annual costs [17, 18, 67, 68]. The above work by the scientists and engineers from petroleum and mining engineering proved it feasible to extract geothermal energy for heating or cooling buildings on the surface. For a proposed geothermal system to be installed in an underground mine, a favourable location would be near a heat source in the surrounding rocks such as a hot sedimentary aquifer. Under these circumstances the expenses for drilling, pipes, and operational maintenance can be considerably reduced. Additionally, the heat loss due to water circulation in pipes can be reduced as well due to the short distances for heat transfer. When the CBM decline curve have reached uneconomic levels we proposed to shut down the valves connecting the geothermal reservoir and use the heat pump infrastructure conventional mode (see section 2) to further cool down the coal reservoir. As seen by the literature review the design life of the conventional geothermal heat harvesting from abandoned coal mines is variable and based on economic factors. Building in the geothermally stimulated CBM recovery, followed by conventional geothermal energy harvesting followed by traditional mining increases the economy significantly. As an important effect both the heat hazards and the risk for coal mine gas outbursts in underground working space can be reduced or eliminated at the same time.

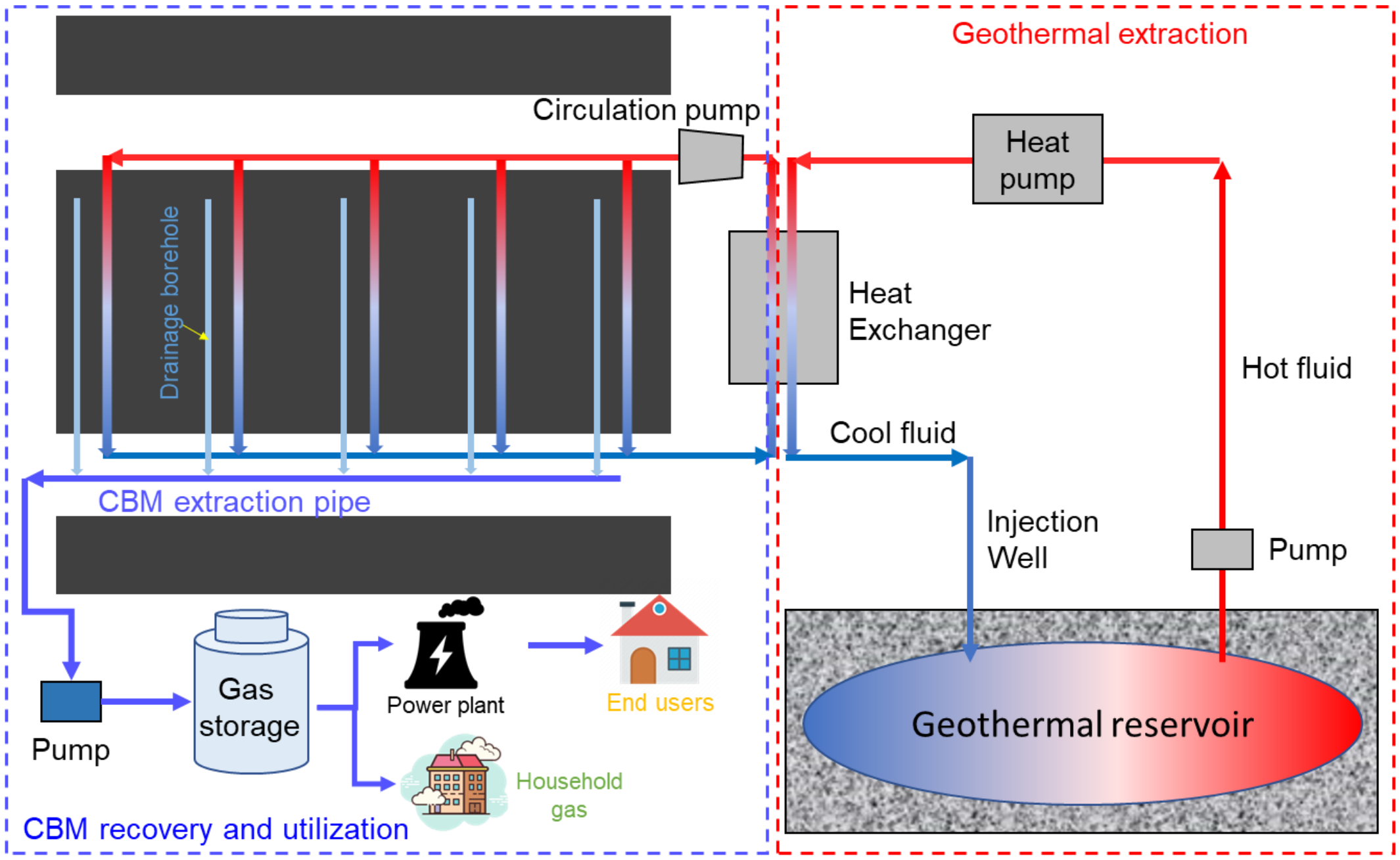

Fig. 9 A schematic diagram of the enhanced CBM by geothermal stimulation.

The temperature dependence of methane ad/desorption has been well documented [53, 56, 58, 60, 64]. It is concluded that the methane desorption increases with the increase of the temperature. The study by our group in Fig. 8 shows that the maximum adsorption of methane drops to approach zero with the increased temperature reaching 200 ℃ [52]. From the experimental data, we can predict that all adsorbed gas can be released from micropores of coal when the temperature of the coal passes a critical value. It means that all embedded methane can be extracted out of the coal seams. However, in reality, the recovered methane is only a small portion of the gas in place. Heat injection, one of many enhanced CBM techniques (including hydraulic fracture, $CO_2$ injection, and heat injection), has been investigated by numerical simulations [69, 70].

Fig. 10 demonstrates the analysis of the data of deep coal mines in China. Fig. 10 (a) and (b) shows the formation temperature gradient, virgin-rock, and working temperature increases over the increase of coal-mine depths. High temperature can also improve the permeability of coal especially those containing high volatile matter. The microwave heating experiment by our team has proven that prolonged heating has a potential for microstructure modification of coal [66].

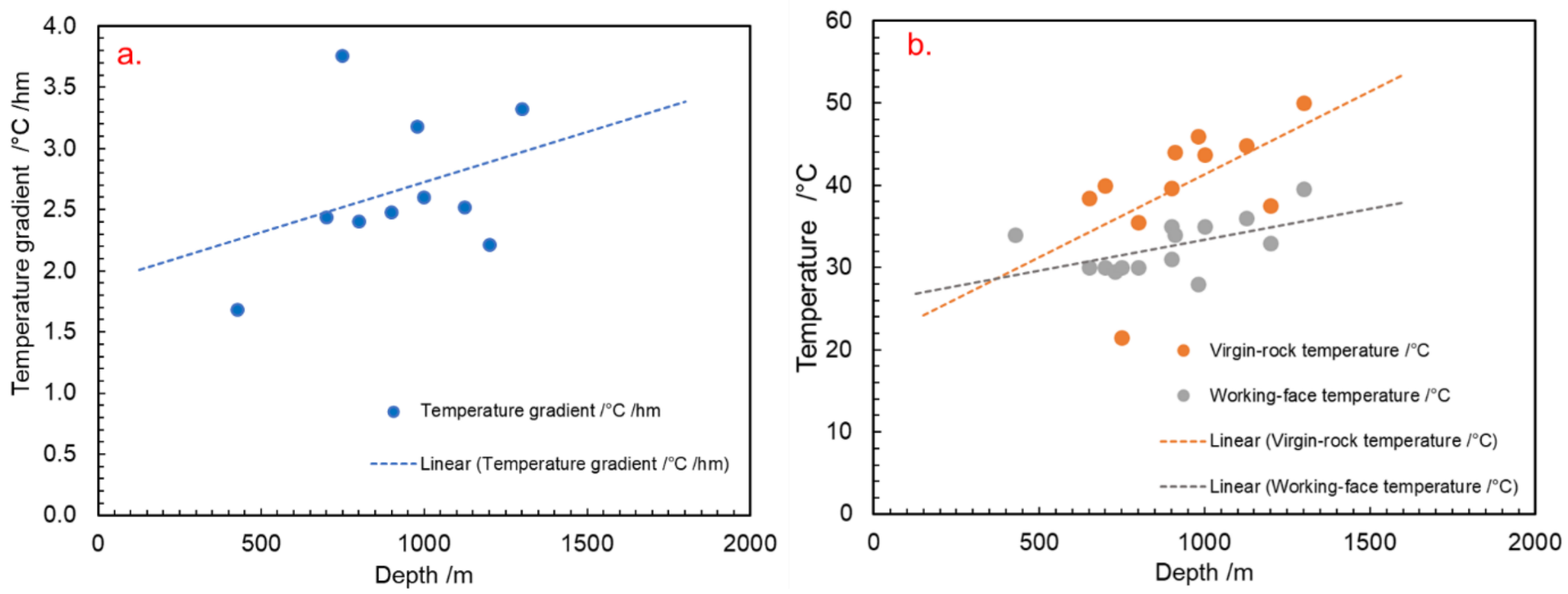

Fig. 10 Evolution of underground temperature over the increase of excavation depth of deep coal mines including (a) temperature gradient of the formation, and (b) virgin-rock temperature, and working-face temperature.

## 5 Future work

The paper only presents the early-stage study of the enhanced CBM recovery by geothermal stimulation in which research status and feasibility analysis have been discussed. The results put forward a positive economic and environmental proposition for the utilization of geothermal energy from the surrounded strata in the deep mining space. Enhancement of CBM production can be one of the potential areas for geothermal application as the methane desorption rate is directly proportional to temperature. Our work will contain two stages including basic theoretical work and field trials. In the first stage, experimental tests, numerical work, and field trials will be sequentially carried out. Currently, little work has been done on thermal fluids, geothermal extraction equipment for the underground environment, and thermal behaviors of methane and coal. This is the theoretical basis of the technique that will determine the whole design of the corresponding system. In the second stage, we intend to collaborate with the mining industry to perform a field trial in a high-temperature coal mine to test the performance of the new system and evaluate the theoretical work.

## 6 Conclusion

This work illustrates the development of geothermal extraction, CBM recovery from deep coal mines in the related fields, and a first feasibility analysis of a newly proposed technique of staggered geothermal stimulation of CBM recovery, geothermal energy use and subsequent mining in the gas depleted and cooled deep coal mine. The following conclusions are made.

(1) Geothermal extraction from underground mines has been well developed and is a robust technique in the field all around the world. It can achieve a large annual energy saving, a reduction of $CO_2$ emission, and monetary savings. It is based on using naturally occurring hot water for geothermal extraction in the deep coal mines.

(2) Heat hazards are serious issues in the coalfields of China, especially in deep coal mines. The data shows the temperature gradient, virgin-rock, and working-face temperature generally increase with the increased excavation depth. High air temperature considerably affects the miners' health and the productivity of mining. The heat from surrounded rock strata in deep coal mines provides a large potential reserve of geothermal resources.

(3) Heat injection in coal seams can improve methane production as the gas desorption rate increases with increasing temperature. The thermal conductivity and microstructures of coal also show a temperature dependence. Our previous work proved that high-temperature heating can enlarge the micropores and improve the permeability of coal.

(4) Therefore, geothermal extraction from deep coal mines is a promising area that can make a great contribution to global energy consumption and eliminate the heat hazards and coal burst induced disasters in the working space. The combination of enhanced CBM by geothermal stimulation is an economical and creative method for cooling underground mines.

## Acknowledgments

This work is supported by the National Key R&D Program of China (2020YFA0711800, 2018YFC0807903) and the Australian Research Council (DP200102517, LP190100990).